\documentclass{emulateapj}
\usepackage{amsmath}
\usepackage{natbib}
\usepackage{epsfig}
\usepackage{bm}

\def\gsim{\;\rlap{\lower 2.5pt
 \hbox{$\sim$}}\raise 1.5pt\hbox{$>$}\;}
\def\lsim{\;\rlap{\lower 2.5pt
   \hbox{$\sim$}}\raise 1.5pt\hbox{$<$}\;}

\begin{document}

\title{An Evolving Entropy Floor in the Intracluster Gas?}
\author{Wenjuan Fang\altaffilmark{1} and Zolt\'an Haiman\altaffilmark{2}}
\altaffiltext{1}{Department of Physics, Columbia University, New
York, NY 10027; wjfang@phys.columbia.edu} \altaffiltext{2}{Department of Astronomy, Columbia
University, New York, NY 10027; zoltan@astro.columbia.edu}

\begin{abstract}
  Non--gravitational processes, such as feedback from galaxies and
  their active nuclei, are believed to have injected excess entropy
  into the intracluster gas, and therefore to have modified the
  density profiles in galaxy clusters during their formation.  Here we
  study a simple model for this so--called preheating scenario, and
  ask (i) whether it can simultaneously explain both global X--ray
  scaling relations and number counts of galaxy clusters, and (ii)
  whether the amount of entropy required evolves with redshift.  We
  adopt a baseline entropy profile that fits recent hydrodynamic
  simulations, modify the hydrostatic equilibrium condition for the
  gas by including $\approx$20$\%$ non--thermal pressure support, and
  add an entropy floor $K_0$ that is allowed to vary with redshift.
  We find that the observed luminosity--temperature ($L-T$) relations
  of low-redshift ($\langle z\rangle=0.05$) HIFLUGCS clusters and
  high-redshift ($\langle z\rangle =0.80$) WARPS clusters are best
  simultaneously reproduced with an evolving entropy floor of
  $K_0(z)=341(1+z)^{-0.83}h^{-1/3} {\rm keV}$ $\rm cm^2$.  If we
  restrict our analysis to the subset of bright ($kT\gsim 3$ keV)
  clusters, we find that the evolving entropy floor can mimic a
  self-similar evolution in the $L-T$ scaling relation. This
  degeneracy with self-similar evolution is, however, broken when
  ($0.5 \lsim kT\lsim 3$ keV) clusters are also included.  The $\sim
  60\%$ entropy increase we find from $z=0.8$ to $z=0.05$ is roughly
  consistent with that expected if the heating is provided by the
  evolving global quasar population.  Using the cosmological
  parameters from the {\it WMAP} 3-year data with $\sigma_8=0.76$, our
  best--fit model underpredicts the number counts of the X-ray galaxy
  clusters compared to those derived from the 158deg$^2$ ROSAT PSPC
  survey. Treating $\sigma_8$ as a free parameter, we find a best--fit
  value of $\sigma_8=0.80\pm 0.02$, in good agreement with the results
  from a recent combined analysis of the Lyman-$\alpha$ forest, 3D
  weak lensing and {\it WMAP} 3-year data.  For the flux--limited
  cluster catalogs, we include an intrinsic scatter in log--luminosity
  at both fixed temperature ($\sigma_{lnL|T}\approx 0.3$) and at fixed
  mass ($\sigma_{lnL|M}\approx 0.6)$, but we find this does not have a
  big effect on our results.
\end{abstract}
\keywords{cosmology: theory: galaxies: clusters: general -- intergalactic medium -- X-rays: galaxies: clusters}

\section{Introduction}

Galaxy clusters, the most massive bound objects in the universe,
provide several methods to constrain cosmological models, for example
through their abundance
\citep[e.g.,][]{Evrard89,HA91,WEF93,ECF96,VL99,Mantz07}, or their
spatial distribution
\citep{schuecker01,RVP02,HH03,BG03,SE03,Linder03}, or both
\citep{schuecker03}. In large future surveys, with tens of thousands
of clusters, percent--level statistical constraints are expected to be
available on dark energy parameters \citep{H&M&H01}, including
constraints on the evolution of its equation of state parameter
$w_a\equiv -dw/da$ \citep{weller2002,weller2003,wang2004}.

In order to fully realize the cosmological potential of large
cluster samples, it is important to understand the cluster
mass-observable relations accurately, at least statistically.  It is
very unlikely that the structure of clusters will be understood from
ab--initio calculations to the level of precision required for the
theoretical uncertainties not to dominate over the exquisite
statistical errors ~\citep[e.g.][]{levine02}. However, in principle,
when multiple observables depend on the same mass, the
mass--observable relation can be accurately determined from the data
itself, simultaneously with cosmological parameters. Several works
have proposed and quantified the constraints from such
`self--calibration'' \citep{majumdar2004,
  wang2004, lima2005}, using parameterized phenomenological relations
for the mass--observable relations (for example, power--law scalings,
or arbitrary evolution in pre-specified redshifts bins).  It has been
argued recently \citep{YHBW06} that even if cluster structure is not
precisely predictable, parameterized physical models can further
improve on such phenomenological self--calibration, especially when
multiple observables (such as X--ray flux and Sunyaev-Zel'dovich [SZ]
decrement) can be predicted from the same physical model
\citep{YHBW06}.  In light of this potential, it is important to fit
physically motivated cluster models to as many cluster observables as
possible; one then hopes that future observations of larger cluster
samples will require further fine--tuning of these models, and, at the
same time, deliver useful cosmological constraints
\citep{OBB05,YHBW06}.

The gravitational potential of clusters is dominated by dark matter,
whose behavior is determined by gravity alone, and is therefore
robustly predictable.  The dark matter profiles of galaxy clusters,
apart from the innermost regions, are indeed well understood from
three--dimensional numerical simulations \citep{NFW97,MGQSL98}, and
are nearly self-similar, as expected.  The physics of gas, on the
other hand, involves complicated non-gravitational processes such as
radiative cooling and star formation, galaxy evolution, and various
forms of feedback.  If these processes were unimportant, the
intracluster gas would trace the self--similar dark matter profile,
and its global properties should obey simple scaling relations
\citep{Kaiser86}.  Specifically, its X-ray luminosity $L$, if
dominated by thermal Bremsstrahlung, as for clusters with temperature
$T>2$ keV, should scale as $L\propto T^2$. This relation is indeed
obeyed by clusters in hydrodynamic simulations without
non-gravitational processes \citep{EMN96,BN98}.  However, the observed
$L-T$ scaling relation is significantly steeper than the self-similar
prediction, closer to $L\propto T^3$ \citep{Markevich98,AE99}. This
demonstrates that the effect of non-gravitational processes on the
intracluster gas is not negligible, even for ``bulk'' observables.

A long--standing proposal for the dominant such non--gravitational
effect is that the intracluster gas is heated by some energy input
(from star formation, supernovae explosion, galactic winds and/or
active galactic nuclei [AGN]), raising the gas to a higher adiabat
before the clusters collapse.  Many authors have investigated the
effect of such a preheating, and have shown that simply imposing a
minimum ``entropy floor'' for the intracluster gas naturally breaks
the self--similarity, and steepens the $L-T$ relation as required by
the data \citep{Kaiser91, EH91, CMT97, TN01, BBLP02, VBBB02}.  The
pre--heating idea is further supported by the discovery of excess
entropy in the inner regions of low--temperature clusters, which
suggests the existence of a universal entropy floor \citep{PCN99,
LDPC00}, and by several other independent lines of evidence \citep[for
a brief summary and a list of references, see, e.g.,][]{BEM01}.

A simple model of pre--heating consists of shifting the entropy
profile by an overall additive constant, representing the cumulative
effect of non--gravitational processes, assumed to be roughly uniform
throughout the gas~\citep[e.g.][]{VBBB02}.  Recent work has tested
this simple model, by comparing its predictions with hydrodynamical
simulations \citep[][hereafter YB07]{YB07}.  The model reproduces the
simulation results very well, but comparisons with observations show
that although it can predict the global X--ray scaling relations, the
model can not reproduce the observed entropy profiles
\citep{PSF03,PA05,PAP06} in detail. This requires the model to be
further developed, but as far as the global properties are concerned,
it appears to be successful, and it is therefore useful to understand
the average properties of the intracluster gas.

In this paper, we adopt this simple preheating model, and focus on
comparisons with both the observed $L-T$ scaling relations in the
redshift range $0\lsim z\lsim 1$, and the observed cumulative number
counts of the X-ray clusters.  A previous study \citep{BEM01}
calculated the impact of preheating on the X--ray scaling relations,
using a sample of 12 simulated clusters, and found a good fit to the
data on local clusters (but has not explicitly compared the expected
evolution to observations, and has not made simultaneous predictions
for the number counts).  Our work is also somewhat similar to a more
recent study by \cite{OBB05}, who present a more detailed physical
model for the intracluster gas, and show that it can reproduce local
X--ray scaling relations (this paper also did not study evolution).

Our goal here is to clarify (i) whether the model can simultaneously
explain both the scaling relations {\it and} number counts of galaxy
clusters, and (ii) whether the amount of entropy required evolves with
redshift.  In comparing our predictions to the $L-T$ scaling relations
and the number counts, we also study the effects of scatter in the
$L-T$ and $L-M$ relations, and the corresponding selection biases that
arise in flux limited survey \citep{NSRE07}.  Our first goal is
motivated by our earlier study \citep{YHBW06}, in which we found that
a similar preheating model, with the entropy adjusted to reproduce
observed X--ray and SZ scaling relations, tends to overpredict the
number counts of bright clusters, even with a relatively low
normalization ($\sigma_8=0.7$) of the power spectrum. A similar
discrepancy was found by \citet[][although they used a higher
normalization, $\sigma_8=0.84$, and suggested that agreement can be
recovered by lowering this value]{OBB05} .

The rest of this paper is organized as the follows. In
\S~\ref{sec:phmodel}, we describe in detail the formalism to
implement the preheating model. In \S~\ref{sec:LTscaling}, we
compare the predicted $L-T$ scaling relations to observations, and
find the best--fit entropy level $K_0$ at two different redshifts.
In \S~\ref{sec:counts}, we further test the model by comparing
predictions for the number counts with observations.  In
\S~\ref{sec:scatter}, we then study the effect of intrinsic scatters
and the corresponding selection effects in flux-limited cluster
surveys. In \S~\ref{sec:discussion}, we discuss our results, and in
\S~\ref{sec:conclusions}, we offer our conclusions.

\section{Modified Entropy Model of Preheating}
\label{sec:phmodel}

We adopt the terminology from the literature, and refer to the
quantity
\begin{equation}
K = \frac{P}{{\rho_g}^{\gamma}}
\label{eq:Ktheory}
\end{equation}
as ``entropy''. Here $P$ and ${\rho}_g$ are the pressure and density
of the gas, and ${\gamma}$ is the adiabatic index. For an ideal gas,
$K$ is related to the formal thermodynamic entropy per particle $s$ by
$s-s_0 \propto \ln K$, with $s_0$ a constant.  In this paper, the
baseline entropy profile to be modified is adopted from YB07\footnote{
To examine the sensitivity of our conclusions below to the choice of
this baseline profile, we also tried adopting the entropy profile of
gas that traces the DM distribution in an NFW halo.  We have verified
that our main conclusion below, that the entropy floor increases with
cosmic time, still holds in thic case. In particular, following the
procedure in \citet{YHBW06}, but assuming $f_g=0.9$ and 20\%
non--thermal pressure support, we find $K_0$ increases from
$363^{+65}_{-61}$ at $z=0.8$ to $507^{+17}_{-17} h^{-1/3}$ keV $\rm
cm^2$ at $z=0.05$ (these numbers include intrinsic scatter, and are to
be compared with the values obtained in our fiducial model in
\S~\ref{subsec:LTscatter}).}, which is a fit to that of the clusters
in AMR simulations \citep{VKB05} without non--gravitational
processes. The profile is self--similar when expressed as a function
of the gas fraction $f_g$, and normalized by $K_{\rm vir}$,
\begin{equation}
\frac{K(f_g)}{K_{\rm vir}}=0.18+0.2f_g+1.5f_g^2.
\end{equation}
Here $f_g(<r)=M_g(<r)/(f_b M_{\rm vir})$ is the gas mass inside
radius $r$, normalized by the cosmic mass fraction of baryons
($f_b=\Omega_b/\Omega_m$) times the total virial mass of the cluster
$M_{\rm vir}$. We further define $T_{\rm vir}=GM_{\rm vir}\mu
m_p/(2r_{\rm vir})$, which is the temperature of the corresponding
isothermal sphere \citep{VBBB02}. (Throughout this paper, we absorb
$k_B$ into $T$, so temperature is in units of energy.) The mean
molecular weight $\mu=0.59$ is adopted for the intracluster gas,
appropriate for a fully ionized H-He plasma with helium mass
fraction $Y_{\rm He}=0.25$; $m_p$ is the mass of proton, and $r_{\rm
vir}$ is the virial radius. $K_{\rm vir}$ is then calculated by
$K_{\rm vir}=T_{\rm vir}/(f_b \rho_{\rm vir})^{\gamma-1}/({\mu
m_p})$, where $\rho_{\rm vir}$ is the mean density of the cluster
within the virial radius (relates $M_{\rm vir}$ to $r_{\rm vir}$ by
$M_{\rm vir}=\frac{4\pi}{3}\rho_{\rm vir}r_{\rm vir}^3$, see below
for its calculation).

The effect of preheating is then realized by adding a constant $K_0$
to $K(f_g)$,
\begin{equation}
K^{\rm ph}(f_g)=K(f_g)+K_0,
\end{equation}
where the value of $K_0$ can be determined once the amount of energy
injected into the cosmic gas, and the density of the gas at the time
of the injection, is specified.  Convective stability requires the
specific entropy $K$ to be a monotonically increasing function of
radius \citep{VBBB02}, and hence of $f_g$. The above prescription of
preheating may change $f_g$ as a function of $r$, but it does not
change the order of the gas shells.  The entropy profile, together
with the hydrostatic equilibrium and gas mass conservation equations,
\begin{eqnarray}
\label{eq:dpdr}
\frac{dP}{dr}=-\eta \rho_g \frac{GM_{\rm tot}(<r)}{r^2} \label{eqn:equil}\\
\label{eq:dMgdr}
\frac{dM_g(<r)}{dr}=4\pi r^2\rho_g
\end{eqnarray}
can be used to solve for the pressure and density distribution of the
intracluster gas. Combined with the equation of state for ideal gases,
the temperature profile of the gas also follows from the solutions. In
equation~(\ref{eqn:equil}), $M_{\rm tot}(<r)= M_{\rm DM}(<
r)+M_g(<r)$. The dark matter profile $M_{\rm DM}(<r)$ is known, and is
given below. Including $\eta$ allows deviations from strict
hydrostatic equilibrium. Here we adopt $\eta=0.8$, the value YB07 find
in their simulations, suggesting that the remaining support for the
gas is provided by turbulent motions.

The boundary condition for $M_g(<r)$ is naturally chosen to be zero at
the origin (to avoid numerical difficulties, in practice we give $M_g$
a small value at some small finite radius). The pressure at the same
position is found by giving it a trial value and integrating
equations~(\ref{eq:dpdr}-\ref{eq:dMgdr}) until the pressure at $r_{\rm
vir}$ matches the expected momentum flux of infalling gas,
\begin{equation}
P(r_{\rm vir})=\frac{1}{3}f_b \rho_{\rm NFW}(r_{\rm vir})v_{\rm
ff}^2.
\end{equation}
Here we assume the accreting gas is cold \citep{VBBLB03}, and that it
falls freely from the turnaround radius ($r_{\rm ta}$) and is shocked
at the virial radius. We assume $r_{\rm ta}=2r_{\rm vir}$, so that the
free-fall velocity $v_{\rm ff}$ from $r_{\rm ta}$ to $r_{\rm vir}$ is
given by $v_{\rm ff}^2=GM_{\rm vir}/r_{\rm vir}(=2T_{\rm vir}/\mu
m_p)$.  The postshock gas density is $f_b \rho_{\rm NFW}$ (see below
for the calculation of $\rho_{\rm NFW}$).  Under extreme conditions,
the free-fall kinetic energy is totally transformed into thermal
energy, and the post--shock gas has a pressure as given above; this
value agrees with that adopted in YB07, matching their simulation
results. (Besides the difference in identifying clusters, our boundary
pressure has a numerical factor of $\frac{2}{3}$ compared to theirs of
0.7.) These two boundary conditions are sufficient for solving
equations~(\ref{eq:dpdr}-\ref{eq:dMgdr}). The result is that the gas
fraction $f_g$ within the virial radius is 0.88 without preheating,
and somewhat less when preheating is turned on.

The matter distribution in virialized clusters is well described by
the NFW \citep{NFW97} model as found from N-body Pure CDM
simulations. Adiabatic hydrodynamical simulations without
non-gravitational processes find gas density profiles quite similar to
the NFW shape, except in the central regions \citep{VBBB02}, where the
gas density levels off. When preheating is turned on, the inner gas
density profile becomes even shallower and deviates more from the NFW
shape. Since the gas is subdominant in mass, we neglect its effect on
the distribution of dark matter. (Though it is found that gas will
cause the dark matter halo to be slightly more concentrated; e.g.
\citealt{Lin06}.)  For simplicity, here we assume the dark matter
profile retains the NFW shape, i.e. $\rho_{\rm
DM}(r)=(1-f_{b})\rho_{\rm NFW}(r)$. For a cluster virialized at
redshift z with mass $M_{\rm vir}$, its NFW density profile is given
as,
\begin{equation}
\rho_{\rm NFW}(r)=\frac{\delta_c \rho_c(z) }{(r/r_s)(1+r/r_s)^2}
\end{equation}
where $\rho_c$ is the critical density of the universe, and $\delta_c$
and $r_s$ are parameters determined from the concentration parameter
$c\equiv r_{\rm vir}/r_s$. We neglect the weak dependence of $c$ on
$M_{\rm vir}$ and $z$, and simply adopt a constant $c=5$ in this
paper.  We identify clusters virialized at redshift $z$ as spherical
regions with mean density $\rho_{\rm vir}=\Delta_v \rho_c(z)$, with
$\Delta_v$ given as a fitting formula by \citet[][based on spherical
collapse model]{KSZBP05},
\begin{equation}
\Delta_v=18\pi^2\Omega_m(z)[1+a\Theta(z)^b],
\end{equation}
where $\Theta(z)=\Omega_m^{-1}(z)-1$, $\Omega_m(z)$ is the matter
density normalized by $\rho_c(z)$, and
$a=0.432-2.001(|w(z)|^{0.234}-1),b=0.929-0.222(|w(z)|^{0.727}-1)$,
with $w(z)$ the dark energy equation of state.

\section{Preheating from the L-T Scaling Relations}
\label{sec:LTscaling}

Once the density, temperature and pressure profiles of the
intracluster gas are specified, global properties, such as the X--ray
luminosity, the emission--weighted temperature, and the
Sunyaev-Zel'dovich decrement can be readily calculated. Here we
compare predictions of the modified entropy model for the
luminosity-temperature scaling relations with those inferred from
X--ray observations. This choice is motivated by simplicity and
robustness: the total luminosity ($L$) and temperature ($T$) can be
inferred from observations without referring to a model for the
intracluster gas. Comparisons to relations involving the mass of the
cluster (such as the mass-temperature relation) are somewhat more
direct from a theoretical point of view, but any such comparison
would, in any case, have to re-derive cluster masses, using
information such as the observed X--ray surface brightness or
temperature profiles, and using our own model, for a fair comparison
with the data.  We also emphasize that similar comparisons with SZ
observables will contain valuable additional information
\citep[e.g.][]{MHBB03,YHBW06}, and should be possible soon with
forthcoming data on cluster profiles from the Sunyaev-Zel'dovich Array
(SZA) survey \citep{muchovej07,tony07}.  We postpone such comparisons
to future work.

The X--ray luminosity $L$ of a cluster is calculated as,
\begin{equation}
L=\int dV \int d \nu n_e(r) n_H(r) \Lambda (T(r), \nu)
\end{equation}
where $n_{e}=(1-Y_{\rm He}+\frac{Y_{\rm He}}{2})\frac{\rho_g}{m_p}$ is
the number density of electrons, $n_H=(1-Y_{\rm
He})\frac{\rho_g}{m_p}$ is the number density of protons, and
$\Lambda$ is the cooling function, calculated by a Raymond-Smith
\citep{RS77} code with metallicity $Z=0.3{\rm Z}_{\odot}$. The
integral is done over the cluster volume $V$ and over frequency
$\nu$. The emission--weighted temperature is calculated as,
\begin{equation}
T_{\rm ew}=\frac{\int dV \int d \nu \rho_g^2(r) \Lambda (T(r), \nu)
T(r)}{\int dV \int d \nu \rho_g^2(r)\Lambda (T(r), \nu) }
\end{equation}
The effect of preheating decreases the central density of the gas, but
increases its temperature. The result is a lower luminosity and a
higher $T_{\rm ew}$; the combined effect at fixed $T_{\rm ew}$ is a
decrease in luminosity.

We compare our predictions to two flux--limited samples of X--ray
clusters.  One is the low--redshift HIghest X-ray FLUx Galaxy Cluster
Sample (HIFLUGCS) presented in \citet{RB02}, including 63 clusters
whose mean redshift is $\langle z\rangle=0.05$. The other is the
high-redshift sample from the Wide Angle ROSAT Pointed Survey (WARPS)
used in \citet{MJES06}, including 11 clusters with a mean redshift of
$\langle z\rangle=0.8$. For each individual cluster, we predict its
observed temperature as the one weighted by the bolometric emission,
\footnote{We find the difference of this temperature from that
weighted by the band emission, which is actually observed, is less
than 4$\%$. We neglect this difference.  We also checked the bias of
the emission-weighted temperature when comparing to the observed
spectroscopic temperatures (see \S~\ref{subsec:Tsl} below).}  and
compare the bolometric luminosity, calculated at this temperature
using the preheating model, with the observed value. To quantify the
goodness of fit of this comparison, we define the usual $\chi^2$
statistic,
\begin{equation}
  \chi^2=\sum_{i=1}^{N}\frac{[\log L(T_i,z_i, K_0)-\log L_i]^2}{
  (\frac{\partial \log L}{\partial \log T}
   \vert_{T_i}\sigma_{\log T_i})^2+\sigma_{\log L_i}^2}.\label{eqn:xchi2}
\end{equation}
Here $\sigma_{\log T_i}$ and $\sigma_{\log L_i}$ denote the
observational measurement errors of (the base 10 logarithm of)
temperature and luminosity, i.e. $\sigma_{\log T_i}\equiv (\log e)
\frac{\sigma_{T_i}}{T_i}$, and $\sigma_{\log L_i}$ is defined
analogously (additional, intrinsic scatter in these quantities will be
discussed below).  We take $z_i, T_i, L_i, \sigma_{T_i}, \sigma_{L_i}$
directly from the published observational data, and $L$ and
$\frac{\partial \log L}{\partial \log T}$ are calculated from the
preheating model.  We fix the parameters of the background cosmology,
adopting the flat $\Lambda$CDM model with the best-fit values from the
{\it WMAP} 3-year results \citep{WMAP3}, i.e. ($h$, $\Omega_mh^2$,
$\Omega_b h^2$, $\sigma_8$, $n_s$) = (0.73, 0.13, 0.022, 0.76, 0.96)

Therefore, in this comparison, $K_0$ is the only free parameter to be
determined by the fit (allowing variations in the cosmological
parameters will be discussed below).  We quote the best--fit entropy
floor value by multiplying $K$, as defined in
equation~(\ref{eq:Ktheory}) above, by a constant factor of $(\mu
m_p)^{\gamma}(n/n_e)^{\gamma-1}$, with $n=\frac{\rho_g}{\mu
m_p}$. This is equivalent to redefining $K$ as
\begin{equation}
K= \frac{T}{{n_e}^{\gamma-1}}
\label{eq:Kobs}
\end{equation}
which is the definition widely used in the observational literature
\citep[e.g.][]{PCN99,PSF03,PA05}. For $\gamma=\frac{5}{3}$, commonly
used units for the entropy defined above is keV $\rm cm^2$, and 1 keV
$\rm cm^2$ corresponds to ejecting
$0.036(1+\delta_b)^{2/3}(1+z)^2(\frac{\Omega_b h^2}{0.022})^{2/3}$ev
per particle to the fully--ionized plasma with overdensity $\delta_b$
and redshfit $z$.

Note that the luminosity inferred from observations is
cosmology--dependent, and since \citet{RB02} and \citet{MJES06} adopt
different values for the cosmological parameters, we re--scale their
quoted luminosity (by the ratio of the luminosity distance--square) to
our fiducial cosmology.

The above procedure, applied to the low-redshift HIFLUGCS clusters,
yields the best-fit entropy floor of $K_0=295^{+5}_{-5}$ $ h^{-1/3}$
keV $\rm cm^2$. We find a total $\chi^2=2293$ for this best fit model,
or a $\chi^2$ per degree of freedom (d.o.f.) of $37$, indicating that
the $L-T$ relation has additional intrinsic scatter (caused, possibly,
by a cluster--to--cluster variation in the entropy floor itself; see
discussion of scatter in \S~\ref{sec:scatter} below).  For the
high-redshift WARPS sample, we find the best fit $K_0$ to be
$172^{+35}_{-33}h^{-1/3}$ keV $\rm cm^2$.  This fit has a total
$\chi^2=7$, or a $\chi^2$ per d.o.f. of $0.7$.

\begin{figure}[t!]
\resizebox{90mm}{!}{\includegraphics{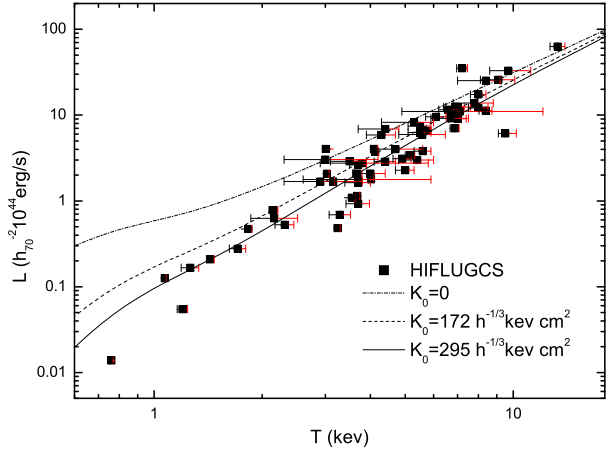}}
\caption{\label{fig:LTlowz} The $L-T$ scaling relation predicted by
the preheating model with the best-fit entropy floor $K_0=295
h^{-1/3}$ keV $\rm cm^2$ at the average redshift $z=0.05$ of the
HIFLUGCS clusters (solid curve), together with data in this sample
from \citet{RB02}, re--scaled to the WMAP 3--year cosmology adopted in
our work.  For reference, we show the $L-T$ relation at $z=0.05$
predicted without an entropy floor ($K_0=0$; dot--dashed curve) and
with the lower entropy inferred from the high--$z$ sample ($K_0=172
h^{-1/3}$ keV $\rm cm^2$; dashed curve; see Figure~\ref{fig:LThighz}).
Measurement errors on $L$ are smaller than the size of the symbols.}
\end{figure}

The $L-T$ scaling relation predicted with the best-fit entropy floor
at the average redshift of the HIFLUGCS clusters, $z=0.05$, is shown
as the solid curve in Figure~\ref{fig:LTlowz}, together with the
re-scaled low--$z$ data from \citet{RB02}.  For reference, the figure
shows the predicted $L-T$ relations without an entropy floor
(dot--dashed curve) and with the lower $K_0$ inferred from the
high-$z$ sample (dashed curve).  The comparison of the data with the
$K_0=0$ curve clearly shows the need for the entropy floor, and the
comparison with the $K_0=172$ $h^{-1/3}$ keV $\rm cm^2$ curve shows
that the observational data, especially of the 1-3 keV clusters,
require that the entropy floor at $z=0.05$ is higher than the
best--fit value at $z=0.8$.

\begin{figure}[t!]
\resizebox{90mm}{!}{\includegraphics{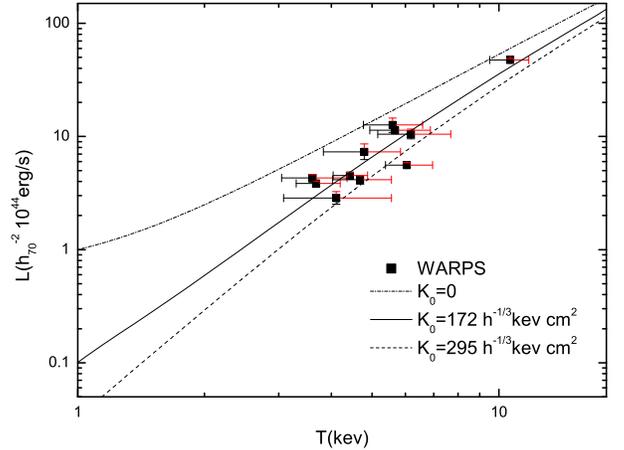}}
\caption{\label{fig:LThighz} The $L-T$ scaling relation predicted by
the preheating model with the best-fit entropy floor $K_0=172
h^{-1/3}$ keV $\rm cm^2$ at the average redshift $z=0.8$ of the
high-redshift WARPS clusters (solid curve), together with the
re-scaled data from \citet{MJES06}.  For reference, we show the $L-T$
relation at $z=0.80$ predicted without an entropy floor ($K_0=0$;
dot--dashed curve) and with the higher entropy inferred from the
low--$z$ sample ($K_0=295 h^{-1/3}$ keV $\rm cm^2$; dashed curve; see
Figure~\ref{fig:LTlowz}).}
\end{figure}

The solid curve in Figure~\ref{fig:LThighz} shows the model prediction
for the $L-T$ scaling relation with the best-fit entropy floor at the
average redshift of the WARPS clusters, $z=0.8$, together with the
re-scaled data from \citet{MJES06}.  For reference, the figure again
shows the predicted $L-T$ relations without an entropy floor
(dot--dashed curve) and with the higher $K_0$ inferred from the
low-$z$ sample (dashed curve).  The comparison of the data with the
$K_0=0$ curve clearly shows the need for the entropy floor at high-$z$
as well, and the comparison with the $K_0=295$ $h^{-1/3}$ keV $\rm
cm^2$ curve shows that the high-$z$ clusters favor an entropy floor
smaller than the best--fit value at low-$z$.

A visual inspection of Figures~\ref{fig:LTlowz} and \ref{fig:LThighz}
(``chi by eye'') indicates that the preheating model of a universal
entropy floor, produced by energy input at an early epoch, can not fit
the scaling relations of the low-redshift and high-redshift clusters
simultaneously.  (We discuss the significance of the detected
evolution quantitatively below, in \S~\ref{subsec:LTscatter} and in
\S~\ref{subsec:diffK}.)  It would be natural, in fact, for the entropy
floor to increase with cosmic time, if the energy input is being
continuously provided by stars and/or AGN.  Parameterizing the entropy
evolution as a power--law in redshift,
\begin{equation}
K_0(z)=K_0(z=0)(1+z)^{-\alpha},
\end{equation}
we can convert the two best-fit values of $K_0$ for the two cluster
samples at $z= 0.05$ and $0.8$ to estimate $K_0(z=0)=310 h^{-1/3}$ keV
$\rm cm^2$ and $\alpha=1$. For reference, this power--law is shown in
Figure~\ref{fig:Kscatter}.

\section{Number Counts of X--ray Clusters}
\label{sec:counts}

The preheating model described above, with the power-law approximation
for the evolution of the entropy floor, can successfully match the
observed $L-T$ scaling relations. This model also predicts a
deterministic relation between cluster mass $M$ and both the
temperature and luminosity. The mass function of dark matter halos is
well understood from both analytic models \citep{PS74,BCEK91,ST99} and
numerical simulations \citep{ST99,Jenkins01}. It is therefore natural
to compare model predictions to observed clusters counts as a function
of either temperature $T$ or luminosity $L$ (or equivalently, flux
$f$).  Here we chose to compare the model predictions to the $\log N-
\log f$ relation derived from the 158 deg$^2$ ROSAT PSPC survey by
\citet{VMFJQH98}.  This sample is ideal for our purposes, since it is
both large and deep enough to provide a good measurement of the counts
to faint fluxes, where the effects of preheating are more pronounced.

We first use the best-fit cosmological model from the {\it WMAP}
3-year results, and calculate $\bar{N}(>$f), the expected surface
number density of clusters whose X-ray fluxes exceed $f$.  The counts
are calculated as
\begin{equation}
\bar{N}(>f)=\int_0^{\infty}
dz\frac{d^2V}{dzd\Omega}(z)\int_{M_{\rm
180}(f,z)}^{\infty}\frac{dn}{dM}(M,z)dM,
\end{equation}
where $d^2V/dzd\Omega$ is the comoving volume element, and
$\frac{dn}{dM}$ is the cluster mass function. In this paper, we use
the fitting formula given by \citet{Jenkins01} for the SO(180) group
finder of dark matter halos. The mass limit $M_{180}(f,z)$ is
determined by first finding the virial mass $M_{\rm vir}(f,z)$ of the
cluster at redshift $z$ that gives a flux $f$; then converting it to
$\rm M_{180}(f,z)$ by extending the NFW profile of this cluster until
the enclosed matter has a mean density of 180 times the background
matter density at that time.

\begin{figure}[t!]
 \resizebox{90mm}{!}{\includegraphics{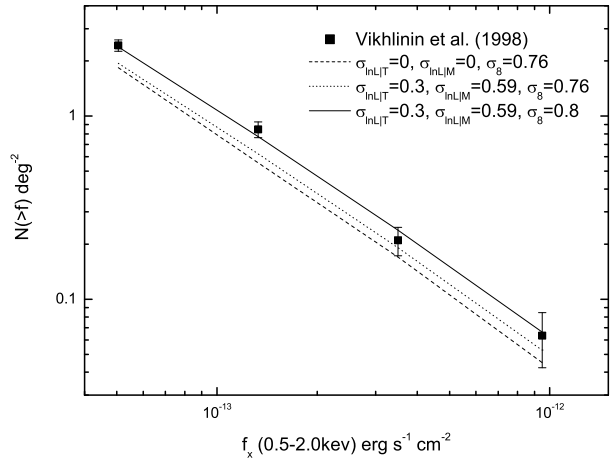}}
 \caption{\label{fig:counts} Cumulative number counts of galaxy
   clusters per deg$^2$ $N(>f)$ as a function of the X-ray flux $f$ in
   the 0.5-2 keV soft X-ray band. Filled squares show data from
   \citet{VMFJQH98}.  The dashed curve shows predictions using the
   {\it WMAP} 3-year cosmology (in particular, $\sigma_8=0.76$), and
   the $L-M$ relation calculated from the preheating model without
   consideration of any intrinsic scatter
   ($\sigma_{lnL|T}=\sigma_{lnL|M}=0$). The dotted curve corresponds
   to the case with intrinsic scatters of $\sigma_{lnL|T}=0.3$ and
   $\sigma_{lnL|M}=0.59$.  The solid curve is similar to the dotted
   curve, except it is calculated with a higher $\sigma_8=0.8$, which
   gives the best agreement with the data when scatter is included.}
\end{figure}

The results are shown as the dashed curve in Figure~\ref{fig:counts},
together with the observational data from \citet{VMFJQH98}.  The
figure shows that the {\it WMAP} 3-year cosmology, together with the
preheating model that fits the $L-T$ scaling relations, underpredicts
the cumulative number counts of X--ray clusters, especially at the low
flux limits.  Considering the sensitivity of the cluster number counts
to $\sigma_8$, it is natural to ask whether the discrepancy can be
resolved by increasing the value of $\sigma_8$ and leaving all other
parameters unchanged (clearly, variations in $\sigma_8$ will not
modify the best--fit $K_0$ inferred from the scaling relations).  We
therefore vary $\sigma_8$, and apply a $\chi^2$ statistic to the
158deg$^2$ ROSAT PSPC data to find its best--fit value. We use
\begin{equation}
\chi^2=\sum_i
\frac{[\bar{N}_i(\sigma_8)-N_i]^2}{\sigma_{N_i}^2+\frac{\bar{N}_i}{A}}\label{eqn:chi2},
\end{equation}
where $i$ labels independent flux bin, $A$ is the survey area. We
include a simple Poisson error (uniform sky coverage at all flux
limits) in the calculation of the variance in addition to the
measurement error. We find the best--fit value of $\sigma_8=0.82\pm
0.02$, which is larger than the {\it WMAP} 3-year best--fit value
$\sigma_8=0.76\pm0.05$ (in the presence of scatter, our best--fit is
reduced to $\sigma_8=0.80\pm 0.02$; see below).

\section{The Effects of Intrinsic Scatter}
\label{sec:scatter}

In the above two sections, we assumed that clusters at redshift $z$
with fixed virial mass $\rm M_{\rm vir}$ have temperatures and
luminosity exactly as predicted by the preheating model. In reality,
deviations from spherical symmetry, as well as cluster--to--cluster
variations in non--adiabatic processes, will lead to non--negligible
scatter in these two quantities.  For flux--limited surveys, such
scatter will cause the observed scaling relations to deviate from the
true ones \citep{SEBSN06,NSRE07}, and the counts to deviate from those
of equivalent mass--limited samples without scatter. To make our
analysis more realistic, it is necessary to take these effects into
account. In this section, we repeat the calculations in the above two
sections, but we include intrinsic scatter, which we model separately
in the $L-T$ and $L-M$ relations.

\subsection{Scatter in the $L-T$ Relation}
\label{subsec:LTscatter}

At a given redshift $z$, the joint probability distribution for $L$
and $T$ of a cluster with fixed $M_{\rm vir}$ may be conveniently
modeled as a bivariate log--normal distribution $P(L,T|M_{\rm vir})$,
with the logarithmic means determined by $M_{\rm vir}$
\citep{NSRE07}. Convolved with the cluster mass function, this can be
used to predict the probability distribution of luminosity for
clusters at fixed temperature $T$.  For a flux--limited sample, the
average and variance of $L$ for these clusters can also be
predicted. Here, since we care only about the final $L-T$ scaling
relation, for simplicity, we assume that $P(L|T)$, the probability
distribution function of $L$ for clusters at fixed $T$ is log--normal,
\begin{equation}
P(L|T)dL=\frac{1}{\sqrt{2\pi}\sigma_{\ln L|T}}
\exp(-\frac{(\ln L-\overline{\ln L})^2}{2\sigma^2_{\ln L|T}})d\ln L.
\label{eq:PLT}
\end{equation}
Given that the log--normal shape of $P(L,T|M)$ is not particularly
well justified to begin with, and that our results are essentially
more sensitive to the width of the $P(L|T)$ distribution than its
detailed shape, we regard this as a sensible approach. The logarithmic
mean $\overline{\ln L}$ in equation~(\ref{eq:PLT}) is taken to be the
logarithm of the luminosity predicted by the preheating model for a
cluster that has temperature $T$ according to the same model. The
scatter $\sigma_{\ln L|T}$ is taken to be a constant. Here we choose
it to be 0.3, which is close to the value$\sim$0.4 expected for
current flux-limited [$f(0.1-2.4\rm keV)$ $\sim 3\times 10^{-12} \,
\rm erg/s/cm^2$] samples. In particular, \citet{NSRE07} derive this
value by assuming a bivariate log--normal distribution of $P(L,T|M)$,
with intrinsic scatters $\sigma_{\ln L|M}=0.59$, $\sigma_{\ln
T|M}=0.1$, power--law relations between the means, and a positive
correlation between $\ln L$ and $\ln T$.

For a flux-limited survey with a threshold $f_{\rm min}$ in the
observer rest frame energy band $[\nu_1,\nu_2]$, the log mean
luminosity of detectable clusters at fixed temperature $T$ is given
by,
\begin{equation}
\langle \ln L\rangle(T)=\overline{\ln L}+\sigma_{\ln L|T}\sqrt{\frac{2}{\pi}}
\frac{\exp(-x_{\rm min}^2)}{{\rm erfc}(x_{\rm min})}\label{eqn:avgLT},
\end{equation}
where erfc is the complimentary error function, $x_{\rm min}=\frac{\ln
L_{\rm min}-\overline{\ln L}}{\sqrt{2}\sigma_{\ln L|T}}$, and $L_{\rm
min}$ is the luminosity corresponding to the flux threshold, $L_{\rm
min}=4\pi d_L^2(z) f_{\rm min}/K(T,z)$. Here $d_L(z)$ is the
luminosity distance, $K$ is the ratio of the X-ray emission in the
energy band $[\nu_1(1+z),\nu_2(1+z)]$ (cluster rest frame) to the
bolometric luminosity, and is calculated by the preheating model for
the same cluster when we calculate $\overline{\ln L}$. Clusters with
luminosity below $L_{\rm min}$ are not included in the average. So,
$\langle \ln L\rangle$ is larger than that for the complete sample
(the so-called Malmquist bias). The variance for the log of the
luminosity for the flux-limited sample can be calculated similarly,
\begin{multline}
\langle(\ln L-\langle\ln L\rangle)^2\rangle(T)=\sigma^2_{\ln L|T}\times\\
\left[1+\frac{2}{\sqrt{\pi}}\frac{x_{\rm min}\exp(-x_{\rm min}^2)}{{\rm erfc}(x_{\rm min})}
-\frac{2}{\pi}\frac{\exp(-2x_{\rm min}^2)}{{\rm erfc}^2(x_{\rm min})}\right].\label{eqn:varLT}
\end{multline}
Note that equations~(\ref{eqn:avgLT}) and (\ref{eqn:varLT}) have
manifestly correct limiting behaviors: in the limit $\rm L_{\rm
min}\rightarrow 0$, $\langle \ln L\rangle\rightarrow \overline{\ln L}$
and $\langle(\ln L-\langle\ln L\rangle)^2\rangle\rightarrow
\sigma^2_{\ln L|T}$; whereas in the limit $L_{\rm
min}\rightarrow\infty$, we have $\langle \ln L\rangle\rightarrow
L_{\rm min}$ and $\langle(\ln L-\langle \ln L\rangle)^2\rangle
\rightarrow \frac{\sigma^2_{\ln L|T}}{x_{\rm min}^2}\rightarrow 0$.

To take into account the above, we modify the calculation of the
$\chi^2$ statistic for the two flux--limited cluster samples.
Specifically, in equation~(\ref{eqn:xchi2}), we replace the average
$\log L(T_i,z_i, K_0)$ by $\langle \ln L\rangle(T_i,z_i,K_0)\times
(\log e)$, and the variance $(\frac{\partial \log L}{\partial \log
T}\vert_{T_i}\sigma_{\log T_i})^2$ by $\langle(\ln L-\langle \ln
L\rangle)^2\rangle(T_i,z)\times (\log e)^2$. Measurement errors in the
temperature may further modify the average and variance of the
$i^{th}$ cluster's luminosity; here, we include this effect
approximately by simply adding a term $\left(\frac{\partial \langle
\ln L\rangle}{\partial \ln T}\sigma_{\log T_i}\right)^2$ to the
intrinsic variance.

With these alterations of the $\chi^2$ statistic, we find that the
best--fit entropy floor $K_0$ for the low--redshift HIFLUGCS clusters
is increased to $327^{+20}_{-19} h^{-1/3}$ keV $\rm cm^2$ (with the
$\chi^2$ per d.o.f of 2.2), and the best-fit $K_0$ for the
high-redshift WARPS clusters is increased to $209^{+66}_{-60}h^{-1/3}$
keV $\rm cm^2$ (with the $\chi^2$ per d.o.f of 0.5).\footnote{Two
clusters in the high-redshift WARPS sample are removed here because
they fall below the nominal flux threshold given in \citet{MJES06}.}
Since Malmquist bias shifts the average luminosity to a larger value,
more entropy is needed to bring the model prediction to agree with the
observations (see \S~\ref{subsec:mbias} for more discussions on this),
but the increase is only $\approx 10-20\%$. More importantly, however,
we see that the significance of the difference in $K_0$ between the
high-- and low--redshift samples is reduced, but remains at the
interesting level of $(327-209)/(\sqrt{19^2+66^2})\approx
1.7\sigma$. (See \S~\ref{subsec:diffK} for more about this.) We find
that the power-law approximated evolution of the entropy floor changes
to $K_0(z)=341(1+z)^{-0.83} h^{-1/3}$ keV $\rm cm^2$.

\subsection{Scatter in the $L-M$ Relation}

In this section, as before, we assume that the bolometric luminosity
$L$ for clusters with virial mass $\rm M_{vir}$ at redshift $z$ has a
log-normal probability distribution,
\begin{equation}
P(L|M_{\rm vir},z)dL=\frac{1}{\sqrt{2\pi}\sigma_{\ln L|M}}
\exp(-\frac{(\ln L-\overline{\ln L})^2}{2\sigma^2_{\ln L|M}})d\ln L.
\end{equation}
The log mean $\overline{\ln L}$ is calculated as the logarithm of the
luminosity predicted for the cluster by the preheating model with the
evolving entropy floor found from \S~\ref{subsec:LTscatter}. The
scatter is taken to be a constant; we adopt the value $\rm \sigma_{\ln
L|M}=0.59$ derived by \citet{SEBSN06} from matching the predicted
cluster counts to the REFLEX survey results \citep{Bohringer04}.

The fraction of clusters with flux above $f$, or luminosity above
$L_{\rm min}$, is then simply
\begin{equation}
P(>f|M_{\rm vir})=\frac{1}{2} {\rm erfc}(x_{\rm min}),
\end{equation}
where $L_{\rm min}$ and $x_{\rm min}$ are calculated as in
\S~\ref{subsec:LTscatter}. Finally, the number counts are given by
\begin{equation}
\bar{N}(>f)=\int
dz\frac{d^2V}{dzd\Omega}(z)\int\frac{dn}{dM}(M,z)P(>f|M_{\rm vir},z)dM.
\end{equation}
Note $M$, the mass of the cluster employed in the mass function, is
defined by an overdensity of 180 of the background matter density,
different from $M_{\rm vir}$. As before, the NFW profile is used to
convert $M_{\rm vir}$ to $M$.  The counts predicted in this model with
scatter are shown as the dotted curve in Figure~\ref{fig:counts}. The
difference from the original calculation, assuming no intrinsic
scatter (dashed curve), is relatively small.  Although a non--zero
$\sigma_{\ln L|M}$, by itself, tends to significantly increase the
number counts, we are also allowing the log mean luminosity
$\overline{\ln L}$ (at fixed $M$) to change. As explained in the
preceding subsection, a non--zero $\sigma_{\ln L|T}$ necessitates more
entropy in order to match the $L-T$ scaling relations, and tends to
reduce $\overline{\ln L}$ (at fixed $T$, and also at fixed $M$), and
hence to decrease the number counts. The combination of these two
effects is that $\bar{N}(>f)$ increases, but only by a relatively
small factor ($\sim 20\%$).

By repeating the analysis as is done at the end of
\S~\ref{sec:counts}, we find that when all other cosmological
parameters are kept fixed at the best-fit values from the {\it WMAP}
3-year results, the preheating model that agrees with the $L-T$
scaling relations at both low and high redshift reduces the best--fit
value of $\sigma_8$ by a small amount, from $0.82^{+0.02}_{-0.02}$ to
$0.80^{+0.02}_{-0.02}$ (see Figure~\ref{fig:counts}). The latter value
still exceeds the best--fit value from the {\it WMAP} 3-year data, but
becomes marginally consistent with their 1$\sigma$ error. We also note
that our best--fit $\sigma_8=0.80$ agrees well with the value found by
\citet{LVHM07} from a combined analysis of Lyman--$\alpha$ forest, 3D
weak lensing and the {\it WMAP} year three data.

\section{Discussion}
\label{sec:discussion}

In this section, we discuss, quantitatively, a range of issues that
should help understand our results and assess their robustness.

\subsection{Significance of the Inferred Entropy Evolution}
\label{subsec:diffK}

Perhaps our most interesting result is the increase in the entropy
floor from the $z\sim0.8$ to the $z\sim0.05$ cluster sample, and
therefore here we discuss the statistical significance of this
difference.  In our analysis above, we have assumed a constant (not
evolving) intrinsic scatter $\sigma_{\ln L|T}$, adapted from the work
of \citet{NSRE07}, resulting in a $\approx 1.7\sigma$ detection for
the difference in the entropy floor values at $z=0.8$ and $z=0.05$
(see \S~\ref{subsec:LTscatter} above). In reality, the measurement
errors of the low-$z$ cluster sample are much smaller than those of
the high-$z$ sample, and the intrinsic scatter can, in fact, be
inferred self--consistently from the $L-T$ relation we fit.  Here we
repeat the analysis in \S~\ref{subsec:LTscatter}, but we allow the
scatter $\sigma_{\ln L|T}$ to vary, and attempt to adjust its value to
find a $\chi^2$ per degree of freedom of unity, for both cluster
samples.  We find that the low-$z$ sample then requires a scatter of
$\sigma_{\ln L|T}=0.49$, which is larger than our adopted value. Using
this larger scatter shifts the best-fit entropy level to
$372^{+37}_{-36}h^{-1/3}$ keV $\rm cm^2$. For the high-z sample, we
find that the measurement errors are so large that the best--fit model
has a $\chi^2$ per d.o.f is less than 1 ($\approx 0.7$) even in the
absence of any intrinsic scatter.  We conclude that the current data
can not yet be used to establish evidence for any intrinsic scatter in
the high--$z$ sample.  The best motivated statistical comparison,
then, is between the best-fit $K_0$ we obtain for the low-$z$ sample
with $\sigma_{\ln L|T}=0.49$, and the best--fit value for the high-$z$
sample obtained with $\sigma_{\ln L|T}=0$ ($172^{+35}_{-33}h^{-1/3}$
keV $\rm cm^2$, see \S~\ref{sec:LTscaling}).  This implies a
significance of the difference between the best--fit values of
$(372-172)/(\sqrt{37^2+35^2})\approx 4\sigma$ (with the best--fit
power--law evolution changing to $K_0(z)=398(1+z)^{-1.43} h^{-1/3}$
keV $\rm cm^2$).  Clearly, better temperature measurements for the
high-$z$ clusters would help determine whether the intrinsic scatter
evolves, which would be important to validate this result.

The entropy floor has a larger impact on the smallest clusters, and
one may wonder to what extent the inferred entropy floor is driven by
the two low--temperature clusters in Figure~\ref{fig:LTlowz} that lie
visibly below the best--fit relation.  When we omit these two clusters
and repeat our analysis with the rest of the HIFLUGCS sample, we find
that the best-fit entropy floor decreases by 7$\%$, from
$295^{+5}_{-5}$ to $273^{+5}_{-5}$ $ h^{-1/3}$ keV $\rm cm^2$ when
ignoring intrinsic scatter in the analysis, by 12$\%$, from
$327^{+20}_{-19}$ to $287^{+22}_{-20} h^{-1/3}$ keV $\rm cm^2$, when
including intrinsic scatter ($\sigma_{\ln L|T}=0.3$), and also by
12$\%$, from $372^{+37}_{-36}$ to $329^{+40}_{-39} h^{-1/3}$ keV $\rm
cm^2$, when including a larger intrinsic scatter ($\sigma_{\ln
L|T}=0.49$). The 4$\sigma$ significance of difference claimed above
now reduces to 3$\sigma$.

\begin{figure}[t!]
\resizebox{90mm}{!}{\includegraphics{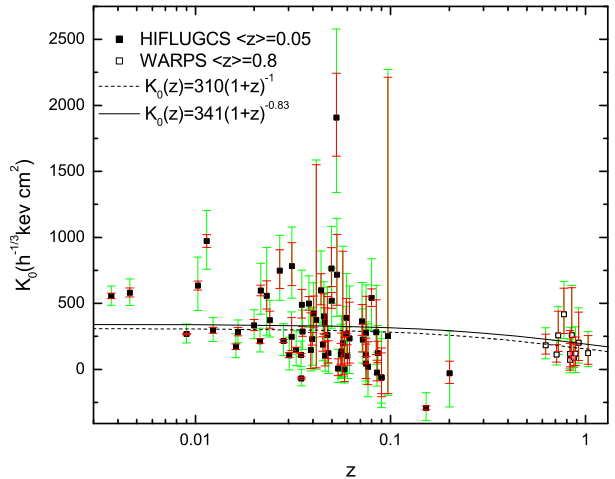}}
\caption{\label{fig:Kscatter} The entropy floor inferred for
individual clusters in the HIFLUGCS and high-redshift WARPS samples,
shown against the cluster's redshift. The narrower (red) error bars
are obtained by allowing the predicted luminosity for the cluster to
vary within the $1\sigma$ regions allowed by observational errors,
while the wider (green) ones are obtained by additionally including a
constant intrinsic scatter in luminosity at fixed temperature
$\sigma_{\ln L|T}=0.3$. The curves are the power--law evolution for
the entropy floor obtained by without taking into account of the
intrinsic scatter (dashed curve) and with the intrinsic scatter of
$\sigma_{\ln L|T}=0.3$ (solid curve).}
\end{figure}

Finally, we use an alternative statistic to assess the significance of
the difference between the high-$z$ and low-$z$ entropy floors. We
derive the entropy floor $K_0$ for each individual cluster in the two
samples by simply setting $L(T_i,z_i, K_0)= L_i$ (following the
notation in \S~\ref{sec:LTscaling}). This results in a range of $K_0$
values, shown by the symbols in Figure~\ref{fig:Kscatter}, which can
be used to construct two separate $K_0$--distributions, for the
high-$z$ and low-$z$ samples.  We then apply a Kolmogorov-Smirnov (KS)
test to the two $K_0$--distributions.  We find $D=0.4$ and a P-value
of 0.07, which makes it unlikely that the two sets of $K_0$ values
were drawn from the same underlying distribution. Unfortunately, this
test remains inconclusive at present, since, as mentioned above, the
observational errors on the temperature are much larger in the
high-$z$ sample than in the low-$z$ sample, and this difference alone
introduces a difference in the inferred $K_0$
distributions. Furthermore, the intrinsic scatter may evolve between
the two redshifts due to reasons unrelated to the entropy floor.
Indeed, this is suggested by the presence of negative $K_0$ values in
the low--$z$ sample, which presumably arises from un-modeled processes
that brighten some clusters' X--ray emission (e.g. cooling cores).  In
order to conclude that the KS tests detects a true evolution (either
in entropy, or in some other process modifying the luminosity
distribution at fixed $T$), we would have to explicitly model the
observational errors, which is not yet warranted, given the large
errors in the high-$z$ sample.

\subsection{Evolution of the X-ray Scaling Relations}

Our analysis requires evolution in the entropy floor, which also
predicts a specific evolution in the $L-T$ scaling relation. In this
section, we compare these evolutions with those derived from
observations in previous work. A particularly relevant study is by
\citet[][hereafter E04]{Ettori04}, which examines the evolution of the
entropy $K$ inferred from the X--ray scaling relation, with $K$
measured at $0.1R_{200}$. They find the entropy $K$ at fixed
temperature evolves as $(1+z)^{0.3}/E_z^{4/3}$, corresponding, from
$z=0.8$ to $z=0.05$, to a 50\% increase, which appears, naively, to be
in good agreement with our finding. However, we caution that E04
measure $K$ at $0.1R_{200}$, which may include a contribution from
gravitational shock heating, especially in higher--mass clusters, and
therefore will not correspond directly to the preheating entropy we
derive (furthermore, the $\beta$--model density profile assumed in E04
differs from the one in our model).

\begin{figure}[t!]
 \resizebox{90mm}{!}{\includegraphics{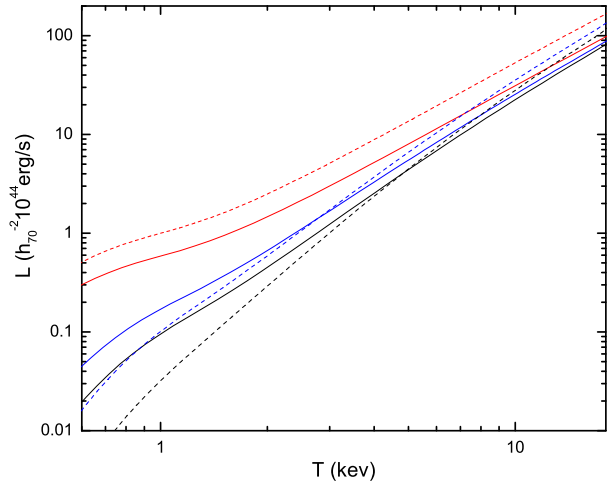}}
 \caption{\label{fig:LTevolution} $L-T$ scaling relations predicted by
 the preheating model at different redshifts and different entropy
 levels.  Solid lines are at redshift $z=0.05$, and dashed lines are
 at redshift $z=0.8$. In both set of lines, from bottom to up the
 entropy floors are set at 295, 172, 0 $h^{-1/3}$ keV cm$^2$. The
 curves are reproduced from Figures~\ref{fig:LTlowz} and
 \ref{fig:LThighz}.}
\end{figure}

Our high-$z$ sample is taken from \citet{MJES06}, which also analyzed
the evolution of the $L-T$ relation, and found that this evolution is
consistent with the expectation in self--similar models (with no
preheating). How can this be reconciled with our results?  We first
note that the high-$z$ sample in \citet{MJES06} includes only clusters
with $T>3$keV, and that their inferred evolution relates to the
normalization of the best--fit power--law relations (whereas our $L-T$
relations are not power--laws). For a clear illustration of how the
two results can be reconciled, we return to our calculations without
intrinsic scatter. In Figure~\ref{fig:LTevolution}, we reproduce the
mean $L-T$ relations from Figure~\ref{fig:LTlowz} and
Figure~\ref{fig:LThighz}, and overlay the six model curves in a single
figure.  Note that the lowest solid curve and the middle dashed curve
are predicted at our best-fit entropy levels for the low-$z$ and
high-$z$ sample, respectively. Comparing these two curves with those
predicted with $K_0=0$, we find our evolving entropy floor predicts a
self-similar-like evolution for the the $L-T$ scaling relations when
$T>3$ keV, in agreement with ~\citet{MJES06}.  This figure also
clearly shows that a constant but non-zero $K_0$ can not mimic a
self-similar-like evolution.

This can be explained by the following: for clusters at the same
redshift, the same entropy level ($K_0$) affects the
low--temperature clusters more than it does the high--temperature
ones, because the latter have larger characteristic
(gravitational-heated) entropy. (This, of course, is well known, and
it is the effect that leads to a larger fractional reduction in the
luminosity for the low--$T$ systems, steepening the $L-T$ scaling
relations.). Similarly, for clusters with the same $T$ but at
different redshifts, a constant entropy level leads to a larger
fractional reduction in the luminosity for the higher redshfit
clusters, because they have larger characteristic density and a
smaller entropy. As a result, maintaining the self-similar-like
evolution requires less entropy at higher redshift.

Provided $T\propto M_{\rm vir}^{2/3}\rho_{\rm vir}(z)^{1/3}$, we have
$K_{\rm vir} \propto T\rho_{\rm vir}(z)^{-2/3}$; to maintain
self-similar evolution of $L$ at fixed $T$, we would need $K_0\propto
\rho_{\rm vir}(z)^{-2/3}$. Taking $K_0=295$ at $z=0.05$, this requires
$K_0=136$ at $z=0.8$, 20$\%$ smaller than our best-fit value at this
redshift. This indicates that the evolution of our $L-T$ scaling
relation is not exactly self-similar, but a little shallower.  Figure
4 in \citet{MJES06} is indeed consistent with this small deviation
from self--similarity.

\subsection{Bias of the Emission-weighted Temperature}
\label{subsec:Tsl}

In our analysis above, we have compared the predicted
emission-weighted temperature $T_{\rm ew}$ for a cluster to its
observational counterpart. Since the latter is generally obtained by
fitting a thermal model to the observed spectrum, in general the
former is a biased estimator. In particular, \citet{Mazzotta04} have
demonstrated that $T_{\rm ew}$ always overestimates the spectroscopic
temperature if the cluster has a complex multi--temperature thermal
structure.  They proposed alternatively using a so--called
spectroscopic--like temperature $T_{\rm sl}$, which they found to be
within a few percent of the actual spectroscopic temperature, measured
for simulated clusters hotter than 2-3 keV. To quantify how the bias
in $T_{\rm ew}$ affect our results, we adopted the formula for $T_{\rm
sl}$ from \citet{Mazzotta04} and repeated our calculations. We find
$T_{\rm sl}$ is larger than $T_{\rm ew}$ by around 10$\%$.  The result
is that the best-fit entropy level shifts to a higher value: from 327
to 420 $h^{-1/3}$ keV $\rm cm^2$ for the HIFLUGCS sample, and from 209
to 287 $h^{-1/3}$ keV $\rm cm^2$ for the WARPS sample, giving the
evolution of $K_0(z)=436(1+z)^{-0.71}$ $h^{-1/3}$ keV $\rm cm^2$. (The
effects of intrinsic scatter in $\sigma_{\ln L|T}$ and Malmquist bias
are included in these results, as in \S~\ref{subsec:LTscatter}.)

\subsection{Parameter Degeneracies}

An obvious issue, even within the context of the simple model adopted
in our study, is that of parameter degeneracies.  A full
multi--dimensional degeneracy study is left for future work; here we
examine only the variations between parameters that we expect may have
the largest effect on our conclusions.

{\em Overall degeneracy between $\eta$, $K_0$, and $\sigma_8$.}  In
our fiducial model, we have included 20$\%$ non-thermal pressure
support (i.e. $\eta=0.8$).  This choice is motivated by simulations
that reveal turbulent motions in the intracluster gas
\citep{NB99,FKNG05,YB07}.  Including turbulent support in the
analytical model is indeed necessary in order to reproduce in detail
the density and temperature profiles for the intracluster gas in
simulations with preheating \citep{YB07}.  There is also direct
observational evidence for turbulence in the Coma cluster
\citep{SFMBB04}.  In addition to turbulence, however, relativistic
particles accelerated by cosmic shocks or other mechanisms can provide
further pressure support for the intracluster gas
\citep{Miniati05}. In order to account for the possibility of such an
additional pressure component, we repeated the analysis of the
previous sections, but changed the value of $\eta$ from $0.8$ to
$\eta=0.7$. This new calculation serves, more generally, to quantify
the impact of uncertainty in the non--thermal pressure component on
our result.

We find that more non--thermal pressure support decreases both the
density and the temperature for a cluster at fixed mass, and decreases
both its $L$ and $T_{\rm ew}$. However, at a fixed $T_{\rm ew}$, we
find that $L$ is slightly increased.  As a result, in order to
reproduce the observed $L-T$ scaling relations, more entropy is needed
(both at low and high redshift). We find the best--fit evolving
entropy floor is changed to $K_0(z)=381(1+z)^{-0.84} h^{-1/3}$ keV
$\rm cm^2$. After preheating is included, keeping the {\it WMAP}
3-year cosmological parameters fixed, the model under--predicts the
number counts even more, as a result of the decreased luminosity at
fixed virial mass. Treating $\sigma_8$ as a free parameter, we find
the best-fit value is increased to $\sigma_8=0.86$. (Intrinsic
scatters are included in the analysis as in \S~\ref{sec:scatter}.)
According to this analysis, non--thermal pressure support is
degenerate with both the entropy floor and the normalization of the
power spectrum: a 50\% increase in non--thermal pressure results in a
8\% increase in $\sigma_8$ and an $\approx 12\%$ increase in $K_0$
(with virtually no effect on the slope of the entropy-evolution).

\begin{figure}[t!]
 \resizebox{90mm}{!}{\includegraphics{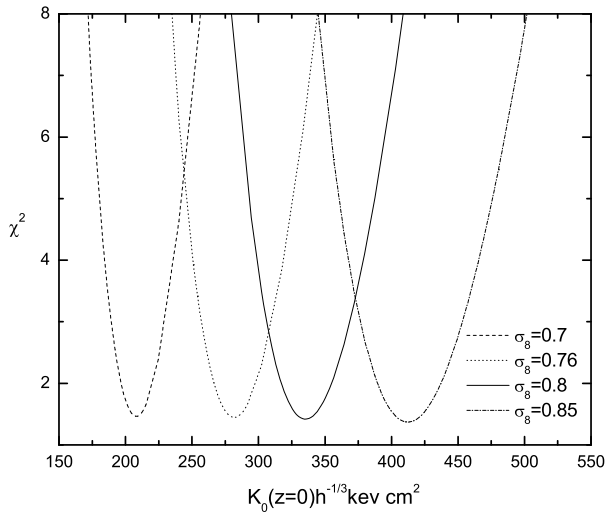}}
 \caption{\label{fig:K0vsSig8} Constraints from the observed X-ray
   cluster number counts \citep{VMFJQH98} on the normalization of the
   evolving entropy floor $K_0(z)=K_0(z=0)(1+z)^{-0.83} h^{-1/3}$ keV
   $\rm cm^2$, for different fixed values of $\sigma_8$.  The
   $y$--axis shows the $\chi^2$, computed from
   equation~(\ref{eqn:chi2}) (Predictions for the counts include
   the intrinsic scatter of $\sigma_{\ln L|M}$).}
\end{figure}

{\em Degeneracy between $K_0$ and $\sigma_8$ from $dN/df$.}  We found
above that if the entropy floor $K_0$ is fitted from the scaling
relations alone, then the best--fit $\sigma_8$ is somewhat higher than
the preferred {\it WMAP} 3--yr value.  It is interesting to quantify
the degeneracy between $K_0$ and $\sigma_8$ from the counts alone --
in particular, to see how large a change in $K_0$ is required if one
insists on the preferred {\it WMAP} 3--yr value of $\sigma_8=0.76$. We
fix the power--law form of the evolution,
$K_0(z)=K_0(z=0)(1+z)^{-0.83}$ (and include a scatter $\sigma_{\ln
L|M}=0.59$, as before), and compute the $\chi^2$ statistic from the
number counts, varying $K_0(z=0)$ and $\sigma_8$ simultaneously. The
results are shown in Figure~\ref{fig:K0vsSig8}.  As this figure
reveals, the best--fit $K_0$ varies monotonically with $\sigma_8$, by
a factor of $\approx 2$ over the range $0.7<\sigma_8<0.85$.  Also, the
best--fit value for $\sigma_8$ from the $L-T$ relation is
significantly discrepant (at the $\approx 2\sigma$ level) from the
central {\it WMAP} 3--yr value of $\sigma_8=0.76$; this discrepancy
can be eliminated by increasing $K_0$ by $\approx 20\%$.

{\em Degeneracy between $\Omega_m$ and $\sigma_8$.}  Cluster number
counts produce a well--known degeneracy between $\Omega_m$ and
$\sigma_8$, approximately of the form $\sigma_8
\Omega_m^{0.5}=$constant for shallow X--ray counts
\citep[e.g.,][]{ECF96,BF98}.

To examine the impact of uncertainty in $\Omega_m$ on our results, we
changed $\Omega_m$ from 0.24 to 0.30 (corresponding to change
$\Omega_mh^2$ from 0.13 to 0.16).  We otherwise fix the {\it WMAP}
3-year cosmological parameters, and re--fit the $L-T$ scaling
relations. We find that the best--fit evolving entropy floor is
decreased significantly, by $\sim 40\%$, to $K_0(z)=194(1+z)^{-0.72}
h^{-1/3}$ keV $\rm cm^2$. This can be understood easily: increasing
$\Omega_m$ decreases the cosmic baryon fraction. For a cluster with
fixed ($M_{\rm vir},z$), the baryon content is therefore decreased.
This reduces its luminosity with the same entropy floor. On the other
hand, the temperature is essentially unchanged. The net result is that
the normalization of the $L-T$ relation is reduced, and less entropy
is needed to bring it into agreement with the observations.

The model with the best--fit entropy floor is then found to {\it over}
predict the X-ray cluster counts. This is mostly due to the increase
in the underlying mass function $dn/dM$, though we also find increased
detection probability for the low mass clusters at a given flux limit,
which may be caused by increased mean luminosity, reduced luminosity
distance, etc.  Allowing $\sigma_8$ to vary, we find the best--fit
value of $\sigma_8=0.66$. This value is smaller than $\sigma_8=0.72$,
the value expected from the usual degeneracy $\sigma_8
\Omega_m^{0.5}$.  (Intrinsic scatters are included in the analysis as
we do in \S~\ref{sec:scatter}.)

\subsection{Which Clusters Are Responsible for the Number Counts Constraints?}

\begin{figure}[t!]
 \resizebox{90mm}{!}{\includegraphics{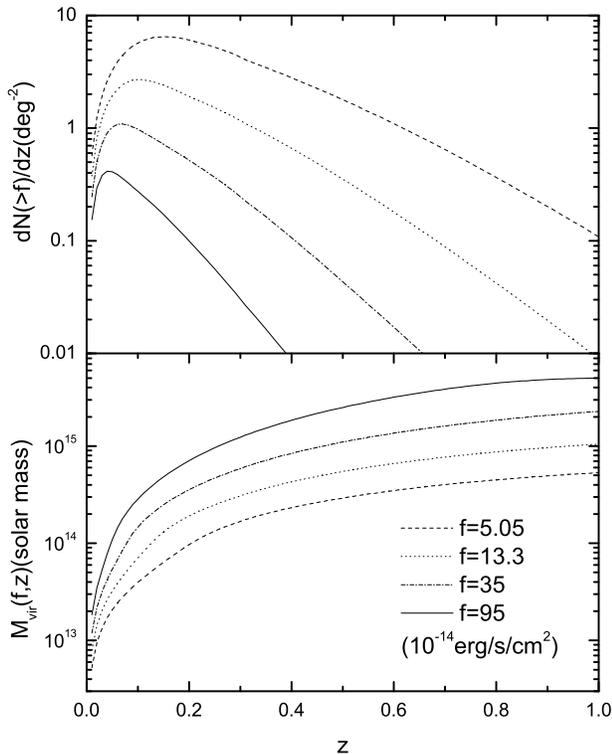}}
 \caption{\label{fig:dndzM} The upper panel shows the redshift
   distribution of the cumulative number density of the X-ray
   clusters, predicted at the flux thresholds of the four data points
   displayed in Figure~\ref{fig:counts}.  Our best--fit preheating
   model is used with $K_0(z)=341(1+z)^{-0.83} h^{-1/3}$ keV $\rm
   cm^2$ and the {\it WMAP} 3-year cosmology, except with
   $\sigma_8=0.8$.  The lower panel shows the mean mass corresponding
   to the four different flux thresholds, as a function of redshift.}
\end{figure}

It is useful to know, within our model, the masses and redshifts of
clusters that dominate the number counts. In Figure~\ref{fig:dndzM},
we show $dN(>f)/dz$ and $M_{\rm vir}(f,z)$ at the four different flux
thresholds we utilized from \citet{VMFJQH98}. The constant intrinsic
scatter of $\sigma_{\ln L|M}$ is adopted in the calculation of
$dN/dz$, so $M_{\rm vir}(f,z)$ is actually the mass of the clusters
that have $50\%$ detection probability. As this figure shows, most of
the clusters are in the range $0.05 \lsim z \lsim 0.15$ and have
masses of $\approx 0.5-2 \times 10^{14}{\rm M_\odot}$. The results
also show that we have included some low mass clusters (few$\times
10^{13} M_\odot$), but the number of these clusters only constitute a
small fraction of the total, e.g. at $f=5.05\times10^{-14}\rm
erg/s/cm^2$, the fraction of clusters with $M_{\rm vir}<5\times
10^{13}M_\odot$ is $\sim13\%$.

\subsection{The Impact of Malmquist Bias on the Entropy Floor}
\label{subsec:mbias}

As an ``academic exercise'', it is useful to assess the impact of
incorporating a flux limit, and the corresponding Malmquist bias, into
our analysis.  For this purpose, we assume that there is an intrinsic
scatter of $\sigma_{\ln L|T}=0.3$ as before, but we do {\it not} apply
any flux limit (this corresponds to setting $x_{\rm min}\rightarrow
-\infty$ in Section~\ref{subsec:LTscatter}). The best-fit entropy
floor is found to be $333 h^{-1/3}$ keV $\rm cm^2$ for the HIFLUGCS
clusters and $174 h^{-1/3}$ keV $\rm cm^2$ for the WARPS clusters (the
two clusters with fluxes below the claimed flux limit are excluded,
for the purpose of fairly comparing with the results that take into
account of the effect of the flux limit).  The entropy evolution is
now $K_0(z)=353(1+z)^{-1.2} h^{-1/3}$ keV $\rm cm^2$.

Compared with the results with no intrinsic scatter, the entropy
levels favored by these two cluster samples both increase. This
increase is caused by the constant intrinsic scatter added to the
denominator in the calculation of the $\chi^2$ analysis, which changes
the relative weight of each cluster (more specifically, reducing the
down--weighting of the (small) clusters that require a higher entropy
floor).

Compared with the results that include the intrinsic scatter and also
apply the survey flux limits, the entropy level for the HIFLUGCS
sample increases a little, while for the WARPS sample, it
decreases. Overall, the impact of the flux limit is surprisingly
modest. One naively expects that the clusters that are most important
for determining the best--fit value for the entropy floor are the
smallest ones, i.e. those just above the detection threshold, which
are most susceptible to bias effects. In particular, a naive
expectation is that this bias will increase the average luminosity,
and will require a larger entropy floor.  It is therefore worth
understanding the relative insensitivity of our results to imposing a
flux limit.

The effects of applying the flux limits have been analyzed in
Section~\ref{subsec:LTscatter}: in addition to increasing the average
luminosity, it also decreases the intrinsic scatter. The former effect
shifts the best-fit entropy to a higher value, while the latter
preferentially increases the value of $\chi^2$ at a larger entropy
floor, and effectively shifts the best-fit entropy level to a lower
value.  Depending on the competition between these two effects, the
net result may be either a larger or a smaller value for the best-fit
entropy floor.  To clarify this competition, we perform an
intermediate calculation, in which the effect of the flux limit is
included only on the average luminosity (i.e.  artificially ignoring
the corresponding reduction in the scatter).  We find the HIFLUGCS
clusters now favor $K_0=411 h^{-1/3}$ keV $\rm cm^2$, and the WARPS
clusters favor $K_0=251 h^{-1/3} $keV $\rm cm^2$.  These value are
much larger than the values obtained by assuming there are no flux
limits, demonstrating that the robustness of the inferred entropy
floor results from the above--mentioned cancelation.  We conclude that
{\it provided the intrinsic scatter is known a--priori (before a flux
limit is applied)}, the effect of Malmquist bias on the inferred
entropy floor is small.

\subsection{Predictions for the SZ Decrement}

\begin{figure}[t!]
\resizebox{90mm}{!}{\includegraphics{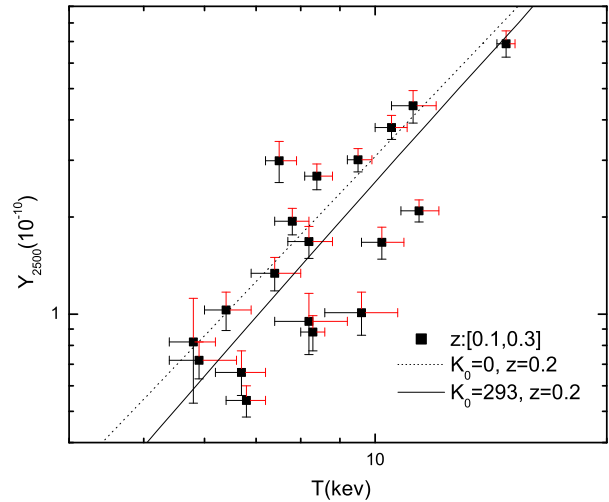}}
\caption{\label{fig:sz} The $Y_{2500}-T$ Sunyaev-Zel'dovich scaling
  relations predicted by the preheating model with the best--fit
  evolving entropy floor given by $K_0(z)=341(1+z)^{-0.83}$ $\rm
  h^{-1/3} keV$ $\rm cm^2$ (solid curve), and without an entropy floor
  (dashed curve) at redshift $z=0.2$. The points with error bars are
  data from \citet{Bonamente07} for clusters within the redshift range
  of [0.1,0.3].}
\end{figure}

As mentioned above, our model fully determines other possible
observables, such as those that can be measured with the SZ effect.
In Figure~\ref{fig:sz}, we plot predictions for the $Y_{2500}-T$
scaling relation, together with the data from \citet{Bonamente07} (see
also \citet{Reese02, MHBB03, Bonamente06, Laroque06} for further
discussions of SZ decrements).  Here $Y_{2500}$ is the integration of
the usual Compton parameter over the solid angle extended by the
cluster within the projected radius of $r_{2500}$ (the radius that
gives a mean enclosed density of 2500 times of the critical density),
and $T$ is the (X--ray) emission--weighted temperature as before. The
solid curve corresponds to our preheating model with the best-fit
evolving entropy floor, and the dashed curve, for reference, shows the
prediction in model without preheating. Both are made at the mean
redshfit of the data $z=0.2$.  A visual inspection of the dashed and
solid curves (``chi by eye'') indicates the data requires preheating,
and that the entropy level we found from the X-ray scaling relations
roughly agrees with the data. A thorough investigation of the SZ
profiles, compared with the X--ray profiles, is likely to yield
interesting new constraints on preheated cluster models
(e.g. \citealt{CLR05}), but we leave this to future work.

\subsection{Moore {\it vs.} NFW Dark Matter  Profiles}

High--resolution numerical simulations suggest that the dark matter
distribution in the central regions of virialized haloes is
significantly steeper than the NFW shape \citep{MGQSL98, KKBP01}. To
examine the dependence of our results on possible variations of the
dark matter profile, here we adopt
\begin{equation}
\rho(r)=\frac{\delta_c \rho_c(z)
}{(r/r_s)^{1.5}(1+r/r_s)^{1.5}}\label{eqn:moore}
\end{equation}
with a fixed concentration parameter $c=4$, and re--compute our
results. We find that the steeper dark matter profile gives a higher
central density and temperature for the intracluster gas, so at a
fixed $M_{\rm vir}$, both L and T are increased, but at a fixed $T$,
the luminosity is decreased. As a result, less entropy is needed for
the preheating model to agree with the observed $L-T$ scaling
relations. We find the favored evolving entropy floor is
$K_0(z)=265(1+z)^{-0.75} h^{-1/3}$ keV $\rm cm^2$. With the {\it WMAP}
3-year best-fit cosmology, the model {\it over}predicts the number
density of the X-ray clusters, and the best-fit $\sigma_8$ is lowered
to 0.74. (Intrinsic scatters are included in the analysis as we do in
\S~\ref{sec:scatter}.)

\begin{figure}[t!]
\resizebox{90mm}{!}{\includegraphics{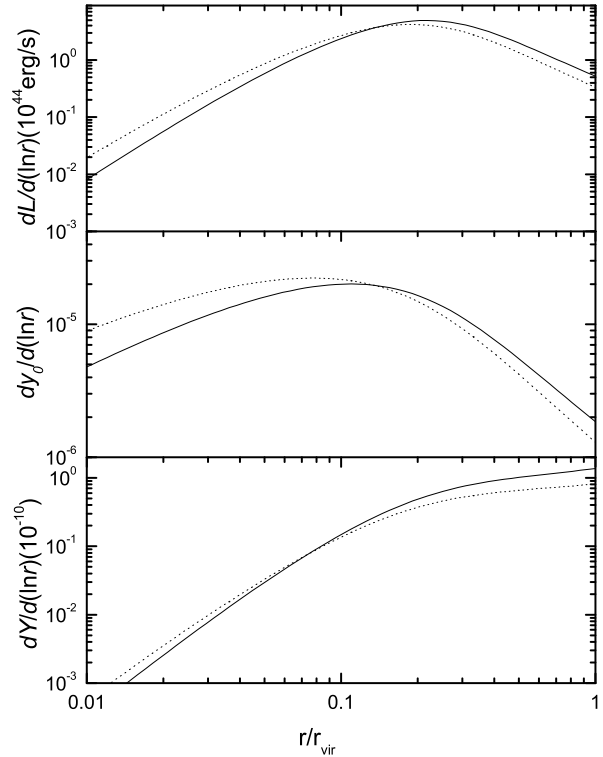}}
\caption{\label{fig:moore} The contributions to X--ray luminosity
  ($L$, upper panel), central SZ decrement ($y_0$, central panel), and
  the integrated SZ decrement ($Y_{\rm vir}$, lower panel) from
  logarithmic radial bins, for both the NFW case (solid curves) and
  Moore et al. case (dotted curves).  The profiles are shown for two
  clusters at $z=0.2$ with the same temperature $T=4.93$keV, predicted
  with $K_0=0$.  The figure demonstrates that $y_0$ is more sensitive
  to the central regions, and is increased by steepening the DM
  profile though the cluster actually has a smaller $M_{\rm vir}$ with
  this profile.  (See discussion in the text.)}
\end{figure}

We also use this steeper dark matter profile to predict the SZ
observables $y_0$ and $Y_{2500}$. We find, similarly to $L$ and $T$,
that at a fixed $M_{\rm vir}$, both $y_0$ and $Y_{2500}$ are
increased; at a fixed $T$, however $Y_{2500}$ is decreased, but $y_0$
is increased. To clarify these changes, we examined two clusters with
the same temperature of $T\sim 5$ keV at $z=0.2$ (predicted by setting
$K_0=0$; we find this requires the cluster to have a mass of $M_{\rm
vir}=9.3 \times10^{14} {\rm M_\odot}$ for the NFW case, and $M_{\rm
vir}=7.3 \times10^{14} {\rm M_\odot}$ for the Moore et
al. case). Since $M_{\rm vir}$ for the Moore et al. case is smaller,
it is understandable that $L$ and $Y_{2500}$ also get smaller (from
$L=8.02\times10^{44}{\rm erg}$ s$^{-1}$ to $L=7.25\times10^{44}{\rm
erg}$ s$^{-1}$, and from $Y_{2500}=5.25\times10^{-11}$ to
$Y_{2500}=4.65\times10^{-11}$).  However, $y_0$ must be more sensitive
to this steeper dark matter profile than the other two observables to
finally get an increase (from $y_0=5.84\times10^{-5}$ to
$y_0=7.18\times10^{-5}$). In Figure~\ref{fig:moore}, we explicitly
show the contributions to $L$, $y_0$ and $Y_{\rm vir}$ (similar to
$Y_{2500}$, except the integration is done within the projected radius
of $r_{\rm vir}$)\footnote{We show $Y_{\rm vir}$ instead of $Y_{2500}$
in order to remove the additional geometrical weighting of different
radial bins. For reference, $Y_{\rm vir}$ decreases from
$1.8\times10^{-10}$ in the NFW case to $1.2\times10^{-10}$ in the
Moore et al. case.} from logarithmic radial bins for both the NFW case
(solid curves) and Moore et al. case (dotted curves). This figure
clearly shows that $y_0$, $L$ and $Y$ is dominated by increasingly
larger logarithmic radius bins. This behavior can be explained by the
fact that the X--ray luminosity and the integrated SZ decrement are
integrations over volume ($\propto r^3)$, whereas the central SZ
decrement is integration over the line of sight ($\propto r$).

\subsection{Comparison with \citet{YHBW06}}

With our best--fit preheating model, adjusted to satisfy the X--ray
scaling relations, and with the {\it WMAP} 3-year cosmology, we found
that the cumulative number counts of the X--ray clusters were {\it
under}predicted.  This is different from the conclusions in earlier
work \citep{YHBW06}, which found an overprediction in a similar model
(\citealt{OBB05} also found an overprediction, using a higher
$\sigma_8$ and a more elaborate cluster structure model).  By
comparing our prediction (without intrinsic scatter) with that of
\citet{YHBW06}, we find that the discrepancy can be attributed to four
differences between our calculation and theirs.  First, we use a
larger value of the entropy floor in the redshift range where the
clusters dominate the number counts, compared with the constant
entropy floor of $194 h^{-1/3}$ keV $\rm cm^2$ adopted by
\citet{YHBW06}. Second, we use the {\it WMAP} 3-year cosmological
model with $\sigma_8=0.76$ instead of the {\it WMAP} 1-year
cosmological model with $\sigma_8=0.7$. Third, in our preheating
model, we use the fitting formula for the baseline entropy profile
from hydrodynamic simulations, which is higher in the central regions
than that adopted by \citet{YHBW06}, and fourth, we also include 20\%
non-thermal pressure support. All of these differences (except for our
larger $\sigma_8$) lead to reductions in the number density, and the
amount of reduction is larger than the increase caused by $\sigma_8$,
leading to a net decrease in the predicted counts.

\subsection{Expected Entropy Evolution}

Since we find evidence for a significant increase in the entropy floor
from the $z\sim0.8$ to the $z\sim0.05$ cluster sample, it is
interesting to ask whether such an evolution is indeed expected if
energy is continuously being injected into the intra--cluster gas.  It
is possible to estimate the entropy history of the IGM from the known
global evolution of AGN and star formation rate.  For example,
\citet{VS99} find that the mean entropy level of the IGM is increasing
with time in both scenarios. In this case, clusters that form at
earlier times will indeed contain gas with a lower entropy
floor. Assuming the resulting entropy floor can be represented by the
background entropy at the formation redshift, we find the entropy
floor for clusters at $z=0.8$ evolves to $z=0.05$ by an increase of a
factor of $\sim2$, according to the calculation of \citet{VS99} for
the AGN heating scenario (see their Figure 2; in the stellar heating
case, the evolution is much steeper, but it is unclear whether stars
can provide the necessary amount of heat).  This increase is
comparable to our findings: $\sim 70\%$ when we assume no intrinsic
scatter, and $\sim 60\%$ when we include an intrinsic scatter. Of
course, this comparison is based on a simple assumption, and the
heating sources in (proto)clusters may be also different from the
global average population. We leave a more serious comparison to
future work.

\section{Conclusions} \label{sec:conclusions}

There is ample evidence that non--gravitational processes, such as
feedback from stars and BHs in galaxies, have injected excess entropy
into the intracluster gas, and therefore have modified its density
profiles.  While in the simplest scenario, the excess entropy is
injected at high redshift, well before clusters actually form, and
results in a universal entropy floor in galaxy clusters. A more
realistic expectation is that the amount of excess entropy evolves
with cosmic epoch, tracking on--going star and BH--formation.

Here we studied a simple model of this preheating scenario, and found
that it can simultaneously explain both global X--ray scaling
relations and number counts of galaxy clusters.  The level of entropy
required between $z=0-1$ is $\sim 200-300\, {\rm keV}$ $\rm cm^2$,
corresponding to $\approx 0.6-0.9 [(1+\delta)/100]^{2/3}[(1+z)/2]^2$
keV per particle if the energy is deposited in gas at overdensity
$\delta$ at redshift $z$.  This overall level of enrichment is in
agreement with previous studies.  Here we find, additionally, evidence
that the entropy floor evolves with redshift, increasing by about
$\sim 60\%$ from $z=0.8$ to $z=0.05$. This fractional increase is in
rough agreement with the evolution expected if the heating rate
follows the global evolution of the AGN.  The normalization
$\sigma_8=0.8$ preferred when X--ray cluster number counts are fit
with our model is somewhat higher than the best--fit value from the
three--year {\it WMAP} data. For a flux--limited cluster catalog, we
also find that including an intrinsic scatter in log--luminosity at
both fixed temperature and at fixed mass does not have a big effect on
the results.

The models presented in this paper should be improved in future work,
refined to fit detailed cluster profiles, in addition to the evolution
of global observables, and allowing a cluster-to-cluster variation of
the level of heating, with possible systematic dependence on cluster
mass, in addition to redshift. It will soon be possible to confront
this type of more detailed modeling with forthcoming SZ
observations. We expect this comparison to securely establish whether
the level of entropy is indeed increasing with cosmic epoch, and to
place further interesting constraints on both cluster structure models
and cosmology.

\acknowledgements

We thank Greg Bryan, Gilbert Holder, Amber Miller, Tony Mroczkowski,
Josh Younger and Caleb Scharf for many useful discussions, and Greg
Bryan, Stefano Ettori, Amber Miller, Josh Younger and Paolo Tozzi for
helpful comments on the draft of this manuscript.  We also thank
Alexey Vikhlinin for providing the X--ray counts
(Figure~\ref{fig:counts}) and Massimiliano Bonamente and Marshall Joy
for providing the SZ data (Figure~\ref{fig:sz}) in electronic form,
with helpful commentary. This work was supported in part by the NSF
grant AST-05-07161, by the Initiatives in Science and Engineering
(ISE) program at Columbia University, and by the the Pol\'anyi Program
of the Hungarian National Office for Research and Technology (NKTH).

\end{document}